# QCD thermodynamics with two flavors of Wilson quarks at $N_t = 6$

Claude Bernard and Michael C. Ogilvie

Department of Physics

Washington University, St. Louis, MO 63130

Thomas A. DeGrand

Department of Physics

University of Colorado

Boulder, CO 80309

Carleton DeTar

Department of Physics

University of Utah

Salt Lake City, UT 84112

Steven Gottlieb and Alex Krasnitz

Department of Physics

Indiana University

Bloomington, IN 47405

R. L. Sugar

Department of Physics

University of California

Santa Barbara, CA 93106-9530

D. Toussaint

Department of Physics

University of Arizona

Tucson, AZ 85721

ABSTRACT: We report on a study of hadron thermodynamics with two flavors of Wilson quarks on $12^3 \times 6$ lattices. We have studied the crossover between the high and low temperature regimes for three values of the hopping parameter, $\kappa = 0.16, 0.17$, and $0.18$. At each of these values of $\kappa$ we have carried out spectrum calculations on $12^3 \times 24$ lattices for two values of the gauge coupling in the vicinity of the crossover in order to set an energy scale for our thermodynamics calculations and to determine the critical value of the gauge coupling for which the pion and quark masses vanish. For $\kappa = 0.17$ and $0.18$ we find coexistence between the high and low temperature regimes over 1,000 simulation time units indicating either that the equilibration time is extremely long or that there is a possibility of a first order phase transition. The pion mass is large at the crossover values of the gauge coupling, but the crossover curve has moved closer to the critical curve along which the pion and quark masses vanish, than it was on lattices with four time slices ($N_t = 4$). In addition, values of the dimensionless quantity $T_c/m_\rho$ are in closer agreement with those for staggered quarks than was the case at $N_t = 4$.

# I. INTRODUCTION

One of the major goals of lattice gauge theory is to determine the properties of strongly interacting matter at very high temperatures. An understanding of the nature of the transition between the low temperature regime of ordinary hadronic matter and the high temperature chiral symmetric regime and of the properties of the high temperature regime is important for the determination of the phase structure of QCD and for the interpretation of heavy ion collisions. It may also cast light on the early development of the universe, which was presumably in the high temperature regime immediately after the big bang. In this paper we report on a study of hadron thermodynamics with two flavors of Wilson quarks on $12^3 \times 6$ lattices.[1]

Most studies of hadron thermodynamics have been carried out with staggered quarks because they retain a remnant of chiral symmetry on the lattice at zero quark mass, a U(1) symmetry. However, flavor symmetry is broken and full chiral symmetry is only restored in the continuum limit. By contrast, chiral symmetry is explicitly broken for the Wilson quark action, and is only expected to be restored in the continuum limit. In view of the different behavior of the two quark formulations on the lattice, it is important to perform thermodynamics calculations with both, and to determine whether they do lead to the same continuum results.

Considerable effort has gone into the study of the thermodynamics of Wilson quarks on lattices with four time slices ($N_t = 4$).[2-6] This work has determined the high temperature "crossover" curve, $6/g_t^2(\kappa)$, in the gauge coupling ($6/g^2$)-hopping parameter ($\kappa$) plane, across which quantities such as the plaquette, the Polyakov loop, and the chiral order parameter vary rapidly. Measurements of the hadron screening lengths in the vicinity of the crossover curve indicate a trend towards chiral symmetry restoration on the high temperature side of the curve.[3,6] Spectrum calculations have been carried out for a number of values of $6/g^2$ and $\kappa$ on the crossover curve. In all cases the pion mass has been found to be large. If the crossover curve intersects the critical curve, $6/g_c^2(\kappa)$, along which the pion and quark masses vanish, it does so only for very strong coupling.[7]

For lattices with four time slices the lattice spacing at the crossover temperature, $a = 1/(4T_c)$, is large, so it is hardly surprising that the thermodynamics of staggered and Wilson quarks are quite different.[6] Recently, extensive calculations of the thermodynamics of two flavors of staggered quarks have been carried out on lattices with six and eight time slices.[8,9] It is therefore of particular importance to push Wilson thermodynamics studies to smaller lattice spacings. A start in that direction was made by the Los Alamos group, which carried out simulations on $8^3 \times 6$ lattices.[4] In this paper we report on a thermodynamics study with two flavors of Wilson quarks on $12^3 \times 6$ lattices. We have determined $6/g_t^2(\kappa)$ for three values of $\kappa$, 0.16, 0.17 and 0.18. For each of these values of $\kappa$, we have performed two spectrum calculations on $12^3 \times 24$ lattices in the neighborhood of the crossover in order to set an energy scale for our calculation and to determine the critical curve, $6/g_c^2(\kappa)$. For $N_t = 6$ the crossover curve has moved closer to the critical curve than it was at $N_t = 4$. However, it is still the case that the pion mass is large at those points along the crossover curve at which we have carried out simulations. For the two largest values of the hopping parameter (smallest values of the quark mass) that we have studied, we found coexistence between the high and low temperature states for runs of over one thousand simulation time units. This result indicates either that there is a very long equilibration time, or that there is a possibility of a first order phase transition. In either case, the behavior is quite different from that found for two flavors of staggered quarks at $N_t = 6$. On the other hand, the spectrum results along the crossover curve yield values of $T_c/m_\rho$ that are closer to those found for staggered quarks on identical lattices than was the case for $N_t = 4$.

In Sec. II we describe our simulation, and in Sec. III we present our detailed results.

## II. The Simulation

We have performed simulations with two flavors of equal mass Wilson quarks using the effective action

$$S_{\text{Eff}} = S_W + \Phi^*(M^\dagger M)^{-1}\Phi, \tag{1}$$

where $S_W$ is the usual Wilson action for the gauge field, $\Phi$ is a complex pseudofermion field, $M$ is the Wilson quark matrix,

$$M_{i,j} = \kappa \sum_\mu \left[(\gamma_\mu + 1)U_{i,\mu}\delta_{i,j-\mu} - (\gamma_\mu - 1)U^\dagger_{i-\mu,\mu}\delta_{i,j+\mu}\right] - \delta_{i,j}, \tag{2}$$

and $U_{i,\mu}$ is the SU(3) matrix associated with the link between lattice points $i$ and $i+\mu$. We make use of the hybrid Monte Carlo algorithm,[10] which requires that we introduce a set of traceless anti-Hermitian matrices, $P_{i,\mu}$, that play the role of momenta conjugate to the $U_{i,\mu}$. The algorithm yields a set of field configurations, $\{P_{i,\mu}, U_{i,\mu}\}$ distributed as $\exp(-H_{\text{Eff}})$, where

$$H_{\text{Eff}} = \sum_{i,\mu} P^2_{i,\mu} + S_{\text{Eff}}. \tag{3}$$

One step in the updating algorithm is the numerical integration of Hamilton's equations for the effective Hamiltonian, $H_{\text{Eff}}$. This requires the introduction of a finite time step, $\Delta\tau$. We have used integration trajectories of one simulation time unit, that is $1/\Delta\tau$ time steps using the normalization of the HEMCGC Collaboration.[11] We accept or reject each of these trajectories with a Metropolis step based on the value of $H_{\text{Eff}}$ at the beginning and end of the trajectory. The fields $P_{i,\mu}$ and $\Phi$ are refreshed at the beginning of each trajectory. For the $P_{i,\mu}$ this merely requires generating a set of Gaussian random numbers, while for the pseudofermion field we use the relation

$$\Phi = MR, \tag{4}$$

where $R$ is a vector of Gaussian random numbers distributed as $\exp(-R^* \cdot R)$.

The numerical integration of Hamilton's equations is performed using the leap frog method. Because this algorithm is time reversal invariant and area preserving, the errors introduced by the finite time step are eliminated by the Metropolis acceptance-rejection step. At each step in the numerical integration we must calculate the vector $(M^\dagger M)^{-1}\Phi$ in order to evaluate the quark contribution to the force. This matrix inversion is carried out using the conjugate gradient algorithm with ILU preconditioning by checkerboards.[12] We use a stopping criterion $10^{-5} > \|r\|/\|\Phi\|$ for the conjugate gradient calculation, where $r$ is the conjugate gradient residual vector and $\|r\|$ and $\|\Phi\|$ are the norms of $r$ and $\Phi$ respectively.

The overwhelming fraction of the computer time in lattice gauge theory calculations is consumed by the conjugate gradient calculations. Since one of these must be made at each time step in the numerical integrations of Hamilton's equations, one would like to maximize $\Delta\tau$. On the other hand, the acceptance probability in the Metropolis step is a decreasing function of $\Delta\tau$, so a compromise must be made. In Tables I and II we enumerate the parameters of our thermodynamics and spectrum runs, including the step size, acceptance probability and number of conjugate gradient iterations required for convergence.

One of the striking features of our runs was the difficulty that the system had in tunneling between the high and low temperature regimes for the two largest values of the hopping parameter that we studied, $\kappa = 0.17$ and $0.18$. The Wilson fermion matrix becomes quite ill-conditioned during the tunneling process leading to a marked increase in the number of conjugate gradient iterations needed to meet the stopping criterion and a decrease in the acceptance probability for the Metropolis steps. We performed some experiments with the hybrid molecular dynamics algorithm, that is with the acceptance-rejection step turned off, but the tunneling rate remained very low. To illustrate this phenomenon we plot in Fig. 1 the real part of the Polyakov loop, the number of conjugate gradient iterations needed for convergence and the acceptance probability for the Metropolis step as a func-

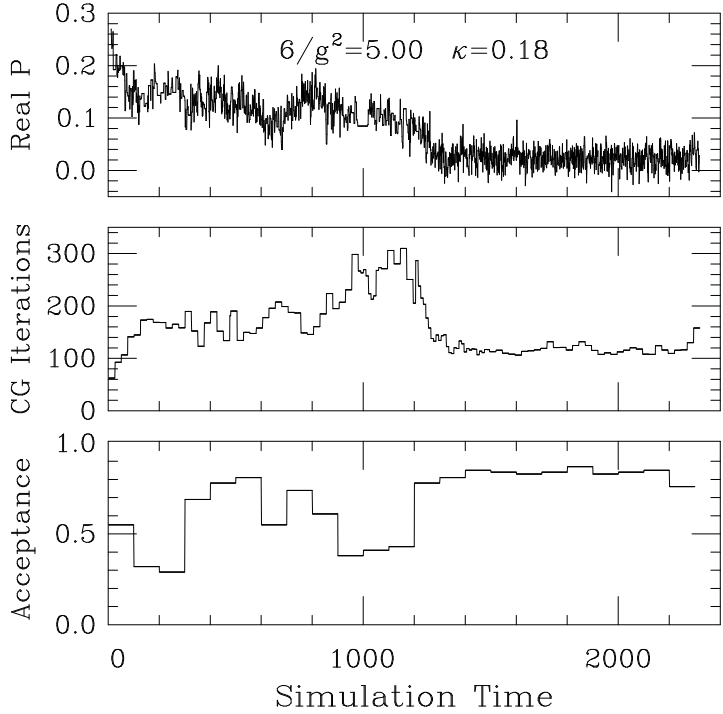

FIGURE 1

The real part of the Polyakov loop, the number of conjugate gradient iterations required for convergence and the acceptance probability for the Metropolis step as a function of simulation time for $6/g^2 = 5.00$ and $\kappa = 0.18$.

tion of simulation time for the run at $6/g^2 = 5.00$ and $\kappa = 0.18$. This run began from a high temperature lattice, whereas the equilibrium state for these parameters is in the low temperature regime. One should particularly note the sharp increase in the number of conjugate gradient iterations and the drop in the acceptance probability during the crossover. (A time step of $\Delta t = 0.018$ was used throughout this run except for the 90 trajectories starting with 1,195. For these trajectories $\Delta t$ was lowered to 0.012 in order to increase the acceptance rate and enable the system to complete the tunneling between the high and low temperature regimes.)

## III. Results

We begin by considering the thermodynamics results on $12^3 \times 6$ lattices. At each hopping parameter that we studied, we performed runs at a range of coupling constants in order to determine the location and nature of the crossover between the high and low temperature regimes. The parameters of these runs are listed in Table I. Thermodynamic quantities such as the real part of the Polyakov loop, the chiral order parameter, $\bar{\psi}\psi$, and the plaquette were measured after each trajectory, that is, every simulation time unit. The data that was collected after the system appeared to reach equilibrium was divided into blocks of sizes that ranged from one to five hundred time units. For each blocking the apparent statistical error was calculated. The errors that we quote are an extrapolation to infinite block size.

For $\kappa = 0.16$ we found a rapid crossover between the high and low temperature regimes, but no evidence for tunneling or metastability. This behavior is consistent with that obtained on lattices with four time slices.[6] We illustrate our $\kappa = 0.16$ results by plotting the real part of the Polyakov loop as a function of $6/g^2$ in Fig. 2. On the other hand, for $\kappa = 0.17$ and $6/g^2 = 5.22$ we found that simulations started from hot and cold lattices had not converged after more than 1,100 simulation time units. We plot the time histories of the real part of the Polyakov loop for these runs in Fig. 3. The higher curve is, of course, from the hot start. We extended this run for over 2,000 time units because it gave some indication that it might evolve to the low temperature regime, but it did not do so. In Fig. 4 we show the real part of the Polyakov loop for the last 1,000 time units collected in the $\kappa = 0.17$ runs at $6/g^2 = 5.20$, 5.22, and 5.23. The runs at $\kappa = 0.18$ show a similar behavior. In Fig. 5 we plot the time histories of the real part of the Polyakov loop for $\kappa = 0.18$ and $6/g^2 = 5.02$ started from hot and cold lattices. In this case there was no indication that either run might move into the opposite regime in the more than 1,000

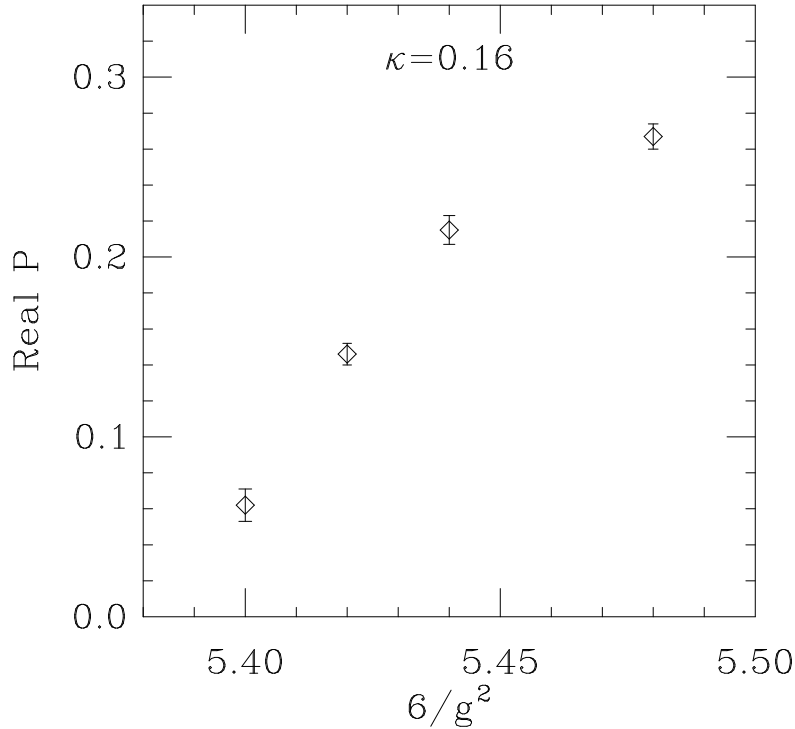

FIGURE 2

The real part of the Polyakov loop as a function of $6/g^2$ for $\kappa = 0.16$.

time units studied. In Fig. 6 we show the time histories of the plaquette for these runs. The large difference in this quantity between the two runs is particularly surprising. In Figs. 7 and 8 we plot the average value of the real part of the Polyakov loop as a function of $6/g^2$ for the runs at $\kappa = 0.17$ and $0.18$. For the points at $\kappa = 0.17$, $6/g^2 = 5.22$, and $\kappa = 0.18$, $6/g^2 = 5.02$ we have calculated averages separately for the runs with hot and cold starts, and included both results on the curves. Whether the results at $\kappa = 0.17$ and $0.18$ are indicative of a first order phase transition or simply indicate that the small time step algorithm used in this study has difficulty in moving the system between the high and low temperature regimes, there is no question that the behavior is very different from that found for two flavors of staggered quarks on $12^3 \times 6$ lattices. The latter exhibit a very smooth crossover between the high and low temperature regimes.[8]

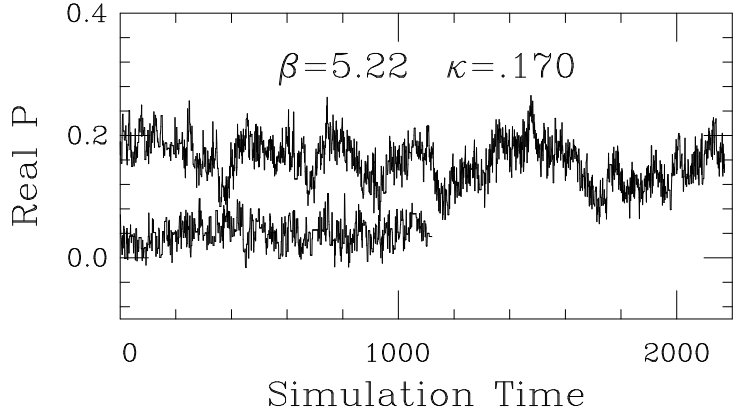

FIGURE 3

Time histories of the real part of the Polyakov for $\kappa = 0.17$ and $6/g^2 = 5.22$. The higher curve is a run started from a hot lattice and the lower curve is a run started from a cold lattice.

Because of the differences between results for $N_t = 4$ and $N_t = 6$, we returned to $N_t = 4$ at one value of the hopping parameter, $\kappa = 0.17$, to check for effects of the spatial size. Earlier work (Ref. 6) on $8^3 \times 4$ lattices found a crossover to high temperature behavior at $6/g^2 = 5.12$, with no strong metastability effects. We found similar behavior on $12^3 \times 4$ lattices. In Fig. 9 we show the time histories of the Polyakov loop for runs at $6/g^2 = 5.11$, 5.12, 5.13, and 5.14, with hot and cold starts at 5.12. This figure may be compared with the $8^3 \times 4$ time histories in Fig. 2 of Ref. 6. This similarity gives us confidence that the effects we are seeing are due to the temperature, or the Euclidean time size of the lattice, rather than the spatial size of the lattice.

For each value of the hopping parameter that we have studied, we have performed two spectrum calculations on $12^3 \times 24$ lattices with values of $6/g^2$ at or near $6/g_t^2$. We have made standard calculations of the $\pi$, $\rho$, $N$, and $\Delta$ masses. These calculations allow us to set a mass scale and to determine the critical value of the coupling at which the pion and quark mass vanish. The parameters for these simulations are given in Table II. For

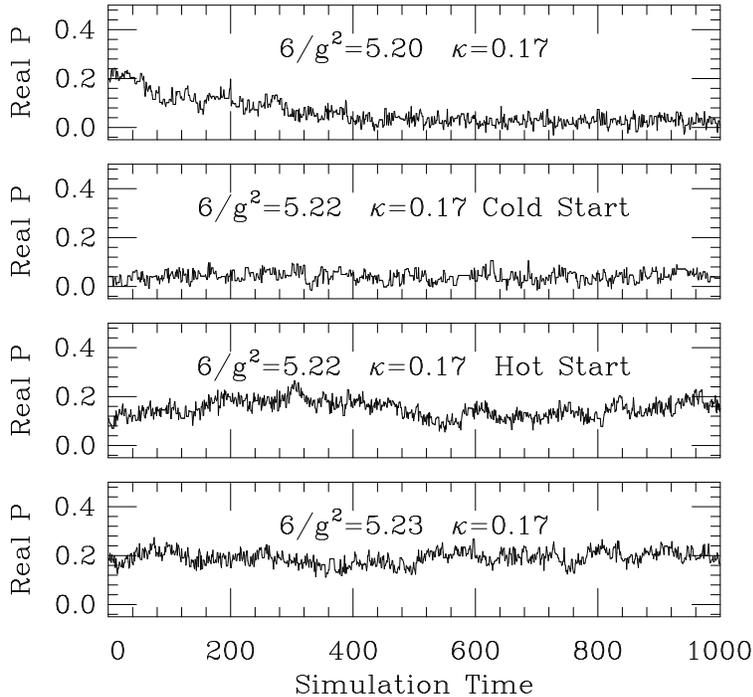

FIGURE 4

The last 1,000 trajectories of the time histories of the real part of the Polyakov loop for $\kappa = 0.17$ and $6/g^2 = 5.20$, 5.22 and 5.23. For 5.22 we show runs started from both hot and cold lattices.

$\kappa = 0.16$ and 0.17 we performed one spectrum calculation precisely at the crossover value of $6/g^2$, and one calculation at a slightly smaller value of $6/g^2$. For $\kappa = 0.18$ we were unable to perform a spectrum calculation at $6/g^2 = 5.02$, because the occurrence of large spikes in the propagators as a function of simulation time made it impossible to obtain statistically meaningful results. We therefore backed off and performed calculations at $6/g^2 = 5.01$ and 4.99. Mass values at the $6/g^2 = 5.02$ were obtained by linear extrapolations from our measured results at 4.99 and 5.01.

In addition to the standard hadron spectrum, we extracted the quark mass from the

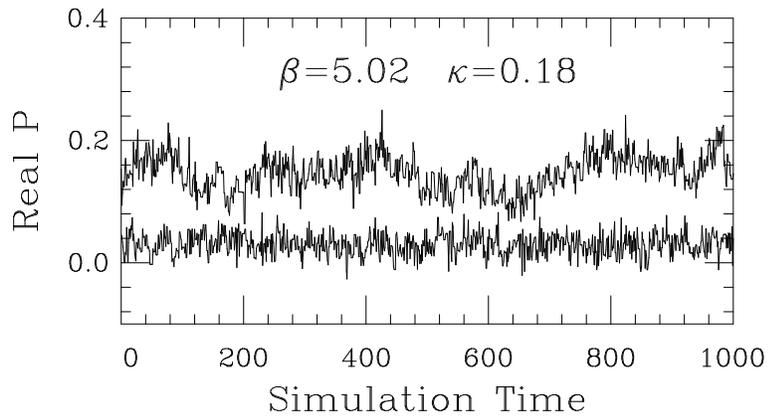

FIGURE 5

Time histories of the real part of the Polyakov loop for $\kappa = 0.18$ and $6/g^2 = 5.02$. The higher curve is a run started from a hot lattice and the lower curve is a run started from a cold lattice.

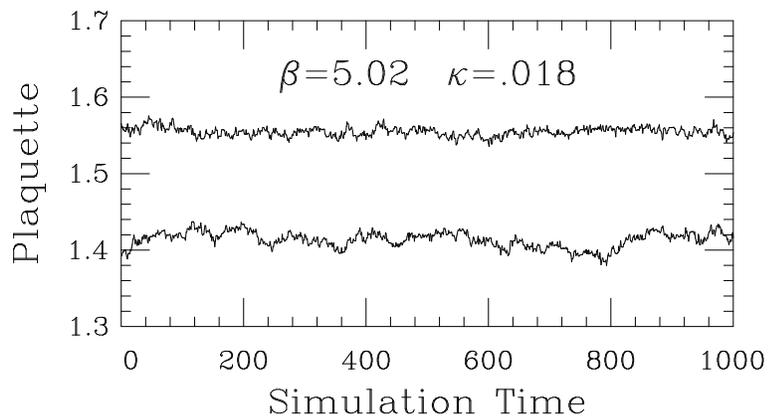

FIGURE 6

Time histories of the plaquette for $\kappa = 0.18$ and $6/g^2 = 5.02$. The higher curve is a run started from a hot lattice and the lower curve is a run started from a cold lattice.

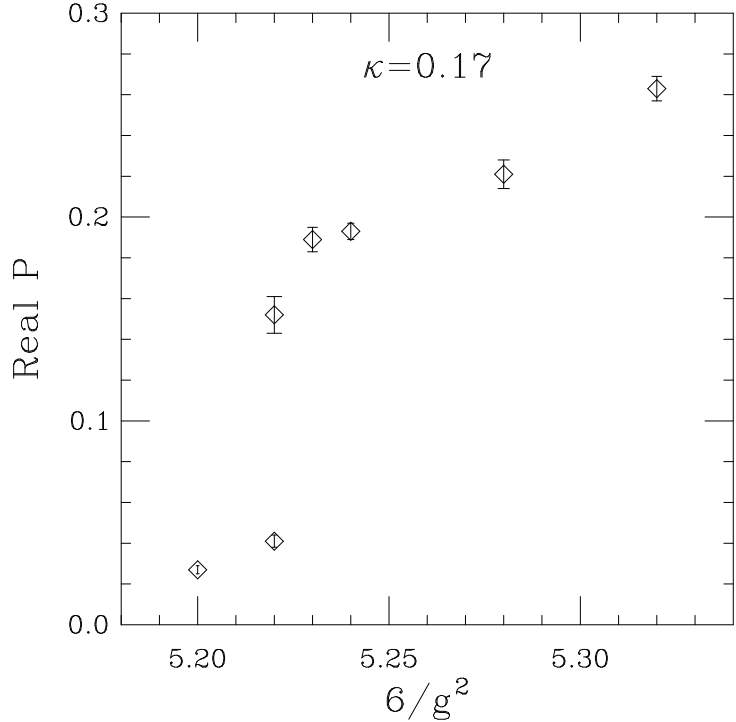

FIGURE 7

The real part of the Polyakov loop as a function of $6/g^2$ for $\kappa = 0.17$.

axial vector current Ward identity through the relations[13]

$$2m_q = \lim_{|x|\to\infty} \frac{\langle \partial_\mu \bar{\psi}(x)\gamma_5\gamma^\mu\psi(x)\bar{\psi}(0)\gamma_5\psi(0)\rangle}{\langle \bar{\psi}(x)\gamma_5\psi(x)\bar{\psi}(0)\gamma_5\psi(0)\rangle}. \qquad (5)$$

We used wall sources and point sinks in these calculations, and extracted the masses by making correlated fits to the propagators. Specifically, we measured two propagators, summed over all spatial points at each time:

$$A(t) = \sum_{\vec{x}} \langle \mathcal{O}_{\text{source}}(0)\bar{\psi}(\vec{x},t)\gamma_5\gamma^0\psi(\vec{x},t)\rangle - (t \to N_t - t), \qquad (6)$$

$$P(t) = \sum_{\vec{x}} \langle \mathcal{O}_{\text{source}}(0)\bar{\psi}(\vec{x},t)\gamma_5\psi(\vec{x},t)\rangle + (t \to N_t - t). \qquad (7)$$

Since the pion is the lightest particle, at large $t$ we only see it. Note that $A(t)$ is antisymmetric in time and $P(t)$ is symmetric. We made a three parameter fit using the full

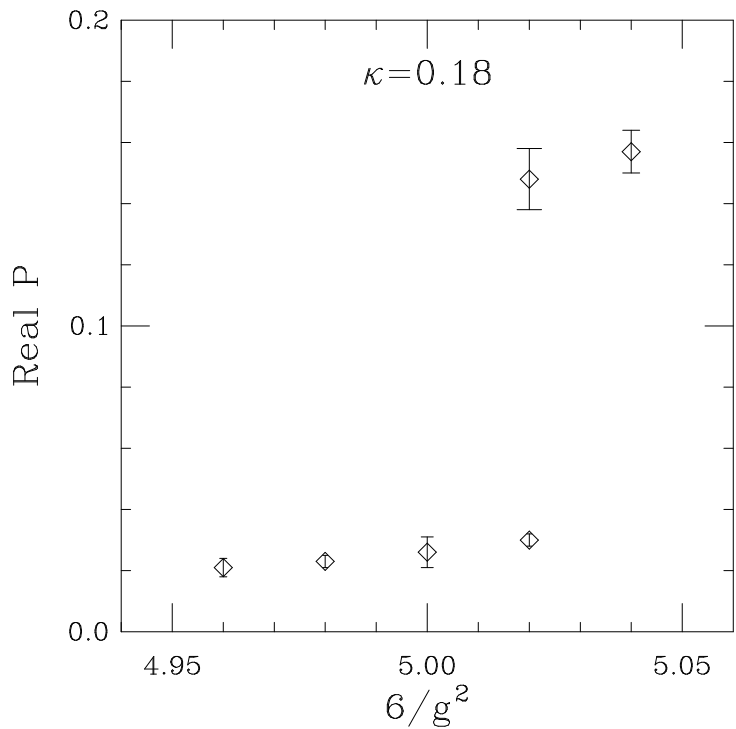

FIGURE 8

The real part of the Polyakov loop as a function of $6/g^2$ for $\kappa = 0.18$.

covariance matrix, with parameters $m_\pi$, $m_q$, and $C$, to the form

$$A(t) = -2Cm_q e^{-m_\pi t} \tag{8}$$

$$P(t) = Cm_\pi e^{-m_\pi t}, \tag{9}$$

so that $\partial A(t)/\partial t = 2m_q P(t)$. This calculation gives an estimate of $m_q$ and its error which takes into account the correlations between the two propagators.

Our spectrum results are tabulated in Tables III–V. The first point to note is that the pion mass does not become small along the crossover curve for the three values of $\kappa$ that we have studied. Just as was the case for $N_t = 4$, in the domain that spectrum calculations have been carried out the crossover and critical curves do not intersect. We have estimated the critical values of the gauge coupling, $6/g_c^2(\kappa)$, for $\kappa = 0.16$, 0.17 and

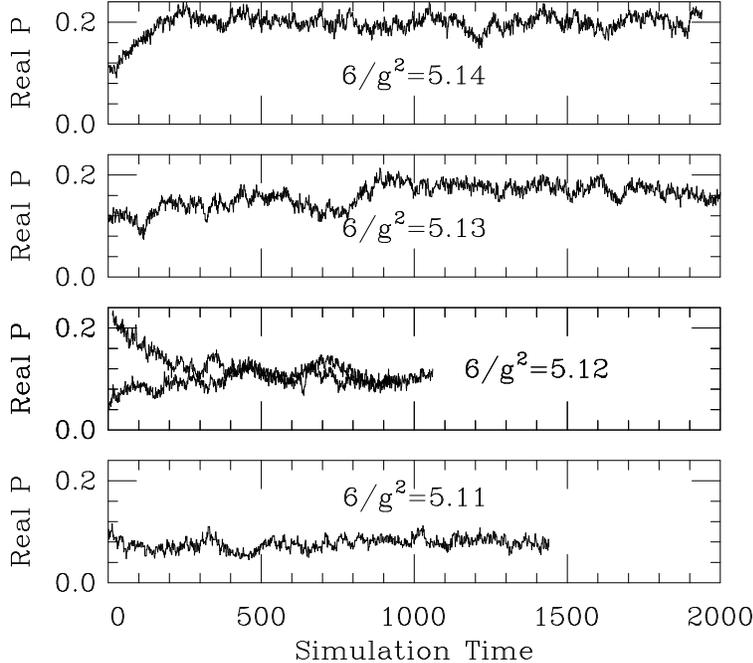

FIGURE 9

The time history of real part the Polyakov loop for $\kappa = 0.17$ and $6/g^2 = 5.11$, 5.12, 5.13 and 5.14 on $12^3 \times 4$ lattices. Runs with hot and cold starts are shown for 5.12.

0.18 by performing linear extrapolations of $m_\pi^2$ and $m_q$ to the points at which they vanish. The results are presented in Table VI. For all three values of $\kappa$ the extrapolations of $m_\pi^2$ and of $m_q$ yield values of the critical coupling which are in very good agreement.

In Fig. 10 we plot our new values of $6/g_t^2(\kappa)$ and $6/g_c^2(\kappa)$. We include in this graph earlier results for $6/g_t^2(\kappa)$ at $N_t = 4$ and 6, and for $6/g_c^2(\kappa)$. It is clear that in going from $N_t = 4$ to 6 the crossover curve has moved closer to the critical curve, but it will be necessary to push to smaller quark masses and lattice spacings in order to study the high temperature regime in the chiral limit.

As stated earlier, one of the objectives of this work was to compare the staggered and

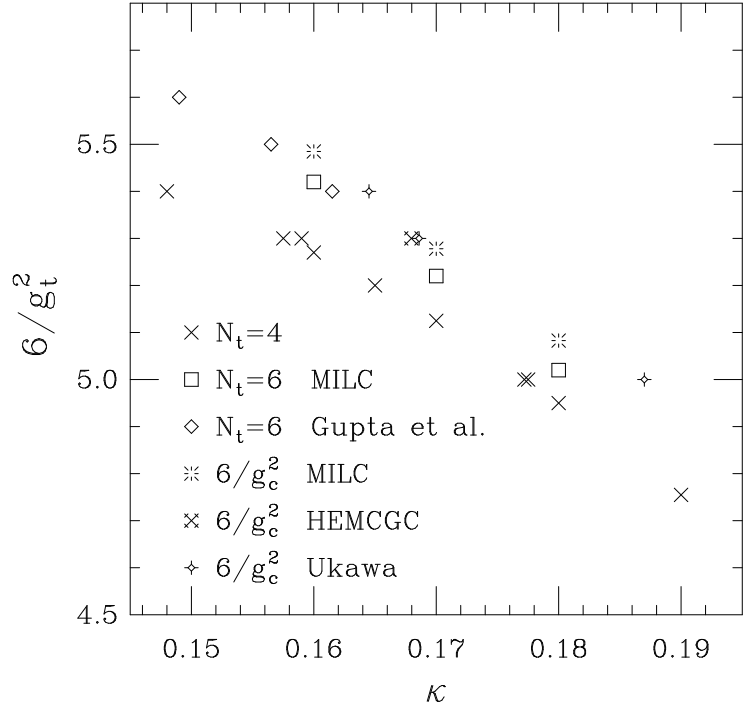

FIGURE 10

Crossover values of the gauge coupling, $6/g_t^2(\kappa)$, as a function of $\kappa$. The squares are our new results on $12^3 \times 6$ lattices, the diamonds are results of the Los Alamos group on $8^3 \times 6$ lattices,[4] and the crosses are a collection of the world data for $N_t = 4$.[3-6] Critical values of the gauge coupling, $6/g_c^2(\kappa)$ are also included. The bursts are our new results on $12^3 \times 24$ lattices, the fancy cross is from the work of HEMCGC Collaboration on $16^3 \times 32$ lattices at $6/g^2 = 5.3$,[14] and the fancy diamonds are Ukawa's results on $6^3 \times 12$ lattices.[3]

Wilson formulations of lattice quarks. To this end we plot in Fig. 11 the dimensionless ratio $T_c/m_\rho$ as a function of $m_\pi/m_\rho$ for values of the gauge coupling and hopping parameter or quark mass on the crossover curve. (In this figure quantities are plotted as calculated in the simulations. No extrapolations are made to the physical limit. The errors in $T_c/m_\rho$ come from the uncertainty in locating the crossover point.) Results are given for both Wilson and staggered quarks at $N_t = 4$ and 6. The dotted line is the physical value of $m_\pi/m_\rho$ to which we would like to extrapolate these results. The Wilson results at $N_t = 6$ are

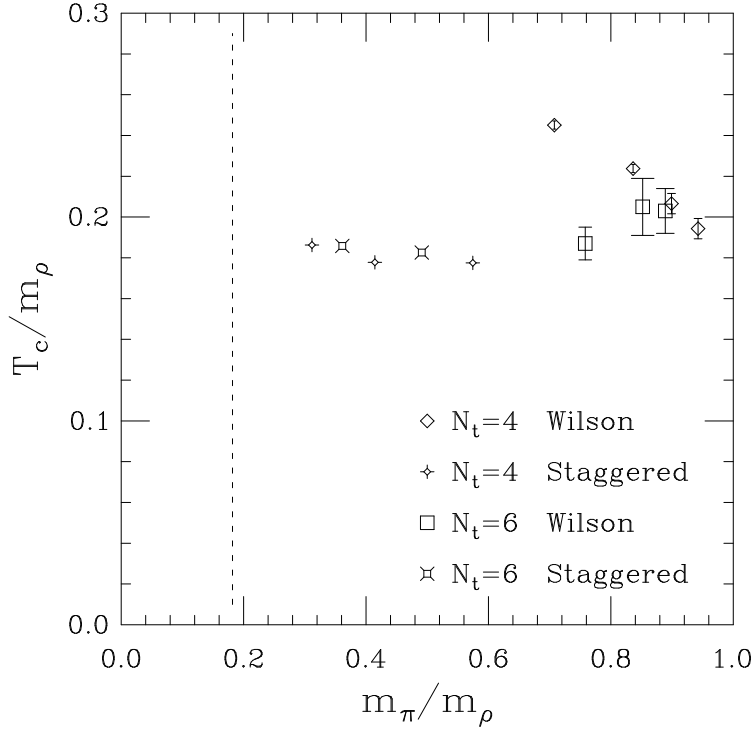

FIGURE 11

The dimensionless ratio $T_c/m_\rho$ as a function of $m_\pi/m_\rho$ for values of the gauge coupling and hopping parameter or quark mass on the crossover curve. The fancy diamonds and fancy squares are the results for staggered quarks at $N_t = 4$ and 6 respectively.[8,15] The ordinary diamonds are the results for Wilson quarks at $N_t = 4$ reported in Ref. 6. The squares are our new results for Wilson quarks at $N_t = 6$.

closer to the staggered ones than was the case at $N_t = 4$. However, it is clearly necessary to push the Wilson simulations to smaller values of $m_\pi/m_\rho$ in order to make a direct comparison. It should also be noted that in view of our comments on the apparent order of the transition at $\kappa = 0.17$ and 0.18 we may, in fact, be observing very different physics than we did in the staggered quark studies. Furthermore, it should be noted that in the Wilson calculations at $N_t = 6$, the coupling is strong and the nucleon to rho mass ratio is large and increasing with $\kappa$.

For staggered quarks it is possible to extrapolate the crossover temperature to the chiral

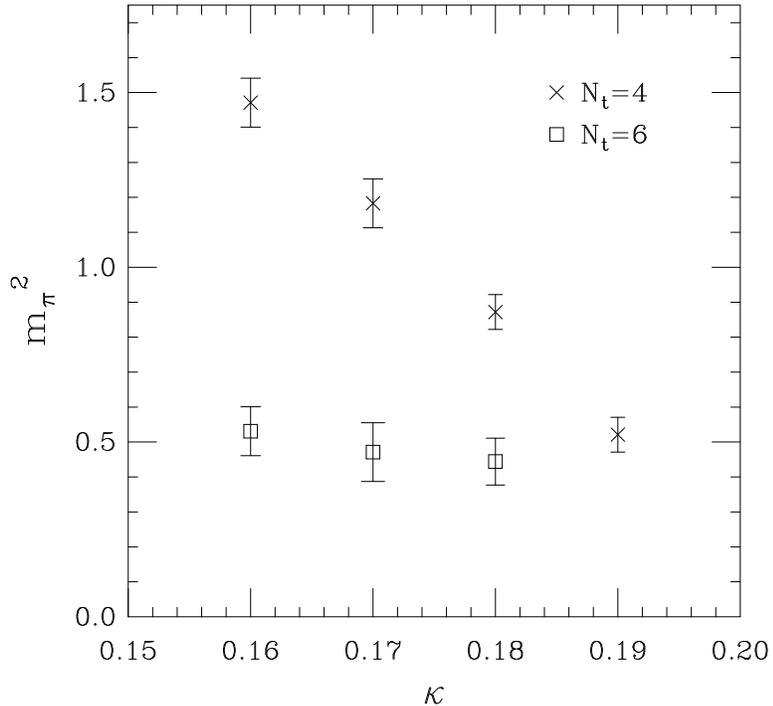

FIGURE 12

$m_\pi^2$ as a function of $\kappa$ for values of the gauge coupling along the crossover curve. The square and crosses are for $N_t = 6$ and 4 respectively.

limit, $m_q = 0$.[15] Calculations for $N_t = 4$, 6 and 8, give results strikingly independent of $N_t$: $T_c \approx 140$ MeV. Similar calculations could be carried out for Wilson quarks. This was done in Ref. 6 for $N_t = 4$ yielding $T_c = 221 \pm 3$ MeV. The quoted error is statistical only. A very considerable extrapolation in $\kappa$ was made, so there is the possibility of a substantial systematic error. Our present $N_t = 6$ data is insufficient to allow us to make to make an extrapolation to $m_\pi^2 = 0$ along the crossover curve. However, the decrease in $T_c/m_\rho$ as $N_t$ is increased from 4 to 6 seen in Fig. 11 indicates that $T_c$ does decrease as the lattice spacing is reduced.

In Fig. 12 we plot $m_\pi^2$ as a function of $\kappa$ for values of $6/g^2$ along the crossover curve. As was the case for $T_c/m_\rho$, the primary source of error is the uncertainty in determining the crossover coupling. Once again, the difference between the $N_t = 6$ and 4 results is manifest.

To summarize, at $\kappa = 0.16$ we found a smooth crossover between the high and low temperature regimes reminiscent of results obtained for staggered quarks on the same size lattice. However, for $\kappa = 0.17$ and 0.18 we found a much sharper crossover than occurs for the lightest mass staggered quarks studied on this lattice. For these values of $\kappa$ the behavior is consistent with a first order phase transition, but may simply be an artifact of the algorithm or of the Wilson hopping matrix. A first order transition would be surprising, particularly since it would not be connected to the deconfining phase transition of pure gauge theory because it is not present for small values of $\kappa$. For the largest hopping parameters studied, the values of $T_c/m_\rho$ are closer to those found for staggered quarks than was the case for $N_t = 4$. However, it is clear that to make a detailed comparison of the two formulations of quarks, one will need to extend the Wilson simulations to smaller lattice spacings and larger hopping parameters. Such calculations will be a very large undertaking.

## Acknowledgements

The calculations reported in this paper were carried out on Intel iPSC/860 hypercubes located at the San Diego Supercomputer Center, the NASA Ames Research Center, and the Superconducting Supercollider Laboratory. We wish to thank all three Centers for their support of our work. This research was supported in part by Department of Energy grants DE–FG02–85ER–40213, DE–AC02–86ER–40253, DE–AC02–84ER–40125, DE–FG02–91ER–40661, DE–AC02–78ER–04915, DE-FG03-90ER40546, and National Science Foundation grants NSF-PHY90-08482, NSF-PHY91-16964, and NSF-PHY91–01853.

# REFERENCES

1. A preliminary report on this work was presented in C. Bernard, T. DeGrand, C. DeTar, S. Gottlieb, A. Krasnitz, M. Ogilvie, R. Sugar, and D. Toussaint, *Nuc. Phys. B (Proc. Suppl.)* **26**, 305, (1992).

2. M. Fukugita, S. Ohta and A. Ukawa, *Phys. Rev. Lett.* **57**, 1974, (1986).

3. A. Ukawa, *Nucl. Phys. B (Proc. Suppl.)* **9**, 463, (1989).

4. R. Gupta, A. Patel, C. Baillie, G. Guralnik, G. Kilcup, and S. Sharpe, *Phys. Rev. D* **40**, 2072, (1989).

5. K. Bitar, A.D. Kennedy, and P. Rossi, *Phys. Lett.* **B234**, 333, (1990).

6. The HEMCGC Collaboration, K.M. Bitar, *et al. Phys. Rev. D* **43**, 2396, (1991).

7. Y. Iwasaki, K. Kananya, S. Sakai and T. Yoshie, University of Tsukuba preprint UTHEP-226.

8. C. Bernard, M.C. Ogilvie, T. DeGrand, C. DeTar, S. Gottlieb, A. Krasnitz, R.L. Sugar and D. Toussaint, *Phys. Rev. D* **45**, 3854, (1992).

9. S. Gottlieb, U. Heller, A.D. Kennedy, J.B. Kogut, A. Krasnitz, W. Liu, R.L. Renken, D.K. Sinclair, R.L. Sugar, D. Toussaint, and K.C. Wang, *Nuclear Physics B (Proc Suppl)* **26**, 308, (1992); and manuscript in preparation.

10. S. Duane, A.D. Kennedy, B.J. Pendleton, and D. Roweth, *Phys. Lett.* **195B**, 216, (1987).

11. The HEMCGC Collaboration, K.M. Bitar, *et al. Phys. Rev. D* **41**, 3794, (1990).

12. T. DeGrand, *Comp. Phys. Comm.* **52**, 161, (1988); T. DeGrand and P. Rossi, *Comp.*

*Phys. Comm.* **60**, 211, (1990).

13. M. Bochicchio, L. Maiani, G. Martinelli, G. Rossi, and M. Testa, *Nucl. Phys.* **B262**, 331, (1985); S. Itoh, Y. Iwasaki, Y. Oyanagi, and T. Yoshie, *Nucl. Phys.* **B274**, 33, (1986); Y. Iwasaki, K. Kanaya, S. Saki and T. Yoshie, *Phys. Rev. Lett.* **67**, 1494, (1991).

14. The HEMCGC Collaboration, K.M. Bitar, *et al.*, to be published.

15. S. Gottlieb, W. Liu, D. Toussaint, R.L. Renken, and R.L. Sugar, *Phys. Rev. Lett.* **59**, 1513, (1987).

Table I. Parameters of the thermodynamics runs on $12^3 \times 6$ lattices. $NT$ is the total number of molecular dynamics time units for the run, $N$ the number of time units retained in calculating ensemble averages after equilibration, $\Delta\tau$ the time step used in integrating the equations of motion, $CG$ the average number of conjugate gradient iterations required for convergence, and $AC$ the acceptance probability for the Metropolis step. The suffixes $h$ and $c$ on values of $6/g^2$ indicate that the run was begun with a hot and cold start respectively.

| $6/g^2$ | $NT$ | $N$ | $\Delta\tau$ | $CG$ | $AC$ |
|---|---|---|---|---|---|
| $\kappa = 0.16$ | | | | | |
| 5.40 | 1,990 | 1,440 | 0.03 | 62 | 0.76 |
| 5.42 | 1,450 | 1,100 | 0.03 | 66 | 0.76 |
| 5.44 | 1,225 | 625 | 0.03 | 63 | 0.74 |
| 5.48 | 1,004 | 654 | 0.03 | 46 | 0.85 |
| $\kappa = 0.17$ | | | | | |
| 5.20 | 1,025 | 525 | 0.035 | 93 | 0.64 |
| 5.22$c$ | 1,120 | 820 | 0.035 | 139 | 0.51 |
| 5.22$h$ | 2,173 | 1,673 | 0.024 | 130 | 0.76 |
| 5.23 | 2,293 | 1,293 | 0.024 | 98 | 0.79 |
| 5.24 | 1,150 | 750 | 0.030 | 98 | 0.70 |
| $\kappa = 0.18$ | | | | | |
| 4.96 | 425 | 325 | 0.018 | 85 | 0.86 |
| 4.98 | 975 | 875 | 0.018 | 96 | 0.86 |
| 5.00 | 2,319 | 1,119 | 0.018 | 134 | 0.83 |
| 5.02$c$ | 1,028 | 928 | 0.018 | 157 | 0.79 |
| 5.02$h$ | 1,050 | 750 | 0.018 | 122 | 0.80 |
| 5.04 | 1,019 | 819 | 0.018 | 93 | 0.57 |

Table II. Parameters of the spectrum runs on $12^3 \times 24$ lattices. $NT$ is the total number of molecular dynamics time units for the run, $N$ the number of time units retained in calculating ensemble averages after equilibration, $\Delta\tau$ the time step used in integrating the equations of motion, $CG$ the average number of conjugate gradient iterations required for convergence, and $AC$ the acceptance probability for the Metropolis step.

| $\kappa$ | $6/g^2$ | $NT$ | $N$ | $\Delta\tau$ | $CG$ | $AC$ |
|---|---|---|---|---|---|---|
| 0.16 | 5.38 | 714 | 400 | 0.02 | 47 | 0.79 |
| 0.16 | 5.41 | 722 | 216 | 0.02 | 62 | 0.75 |
| 0.17 | 5.20 | 632 | 558 | 0.015 | 72 | 0.84 |
| 0.17 | 5.22 | 452 | 226 | 0.015 | 107 | 0.79 |
| 0.18 | 4.99 | 578 | 418 | 0.0115 | 89 | 0.87 |
| 0.18 | 5.01 | 634 | 436 | 0.0115 | 124 | 0.84 |

Table III. Mass spectrum results for $\kappa = 0.16$.

| Particle | $6/g^2 = 5.38$ | $6/g^2 = 5.41$ |
|---|---|---|
| $m_q$ | 0.159(1) | 0.117(2) |
| $m_\pi$ | 0.869(3) | 0.729(4) |
| $m_\rho$ | 0.959(4) | 0.820(7) |
| $m_N$ | 1.55(1) | 1.33(1) |
| $m_\Delta$ | 1.59(1) | 1.38(2) |

Table IV. Mass spectrum results for $\kappa = 0.17$.

| Particle | $6/g^2 = 5.20$ | $6/g^2 = 5.22$ |
|---|---|---|
| $m_q$ | 0.116(1) | 0.088(2) |
| $m_\pi$ | 0.799(3) | 0.686(7) |
| $m_\rho$ | 0.937(5) | 0.81(1) |
| $m_N$ | 1.54(2) | 1.32(1) |
| $m_\Delta$ | 1.60(2) | 1.46(3) |

Table V. Mass spectrum results for $\kappa = 0.18$. The values quoted at $6/g^2 = 5.02$ are extrapolated from the measurements at $6/g^2 = 4.99$ and $5.01$.

| Particle | $6/g^2 = 4.99$ | $6/g^2 = 5.01$ | $6/g^2 = 5.02$ |
|---|---|---|---|
| $m_q$ | 0.102(1) | 0.080(1) | 0.069(2) |
| $m_\pi$ | 0.804(4) | 0.715(5) | 0.666(6) |
| $m_\rho$ | 1.004(5) | 0.93(1) | 0.89(2) |
| $m_N$ | 1.72(1) | 1.63(1) | 1.59(5) |
| $m_\Delta$ | 1.78(2) | — | — |

Table VI. Critical values of $6/g^2$ as a function of $\kappa$. The values labeled $6/g_c^2(m_\pi^2)$ were obtained by performing a linear extrapolation to $m_\pi^2 = 0.0$, while those labeled $6/g_c^2(m_q)$ were obtained by performing a linear extrapolation to $m_q = 0.0$.

| $\kappa$ | $6/g_c^2(m_\pi^2)$ | $6/g_c^2(m_q)$ |
|---|---|---|
| 0.16 | 5.485(3) | 5.494(6) |
| 0.17 | 5.278(7) | 5.283(7) |
| 0.18 | 5.086(6) | 5.083(8) |